\documentclass{article}
\usepackage{spconf,amsmath,graphicx}

\usepackage{enumitem}
\setlist{nosep, leftmargin=14pt}
\usepackage{subcaption}
\usepackage{bm}

\usepackage{amssymb}
\usepackage{mathtools}
\usepackage{amsmath}
\usepackage{mwe} 

\title{WSSAMNet: Weakly Supervised Semantic Attentive Medical Image Registration Network}
\name{Sahar Almahfouz Nasser$^*$, Nikhil Cherian Kurian$^*$, Saqib Shamsi$^  \dagger$, Mohit Meena$^*$, and Amit Sethi$^*$\thanks{Code for the paper can be accessed at: (https://github.com/SaharAlmahfouzNasser/BraTSReg-Challenge)}}
\address{$^*$Indian Institute of Technology Bombay, Mumbai, India.\\
$^\dagger$Whirlpool, Pune, India.}
\begin{document}
%
\maketitle
%
\begin{abstract}
We present WSSAMNet, a weakly supervised method for medical image registration. Ours is a two step method, with the first step being the computation of segmentation masks of the fixed and moving volumes. These masks are then used to attend to the input volume, which are then provided as inputs to a registration network in the second step. The registration network computes the deformation field to perform the alignment between the fixed and the moving volumes. We study the effectiveness of our technique on the BraTSReg challenge \cite{baheti2021brain} data against ANTs and VoxelMorph, where we demonstrate that our method performs competitively.
\end{abstract}
%
%
\section{Introduction}
\label{sec:intro}

Image registration is the process of finding spatial correspondence between two or more images. It has various applications in medical imaging problems such as in multi-modal image fusion or digital subtraction angiography (DSA), where it is used to map structural changes in tumors before and after treatment or identifying tissue atrophy in degenerative diseases. Brain Tumor Sequence Registration (BraTSReg) challenge \cite{baheti2021brain} is the first challenge to address the problem of registering post-treatment follow-up scans of the magnetic resonance imaging (MRI) to the pre-operative MRI scans of patients treated for glioma. The change in the tissue appearance caused by the pathologies makes this registration very challenging even though we are dealing with a single imaging modality. 


In general, registration comprises multiple steps such as feature detection and matching, designing the mapping function, image transformation, and image resampling.
For the registration methods, the detected features should be distinctive features such as closed-boundary regions, edges, contours, line intersections, and corners. However, unlike natural images medical images are often sparse in features; thus the registration methods for medical images, in general, are region-based methods. For feature matching, the registration methods depend on the similarity measures between the descriptors of the detected features, such as cross-correlation, mutual information, cross-power spectrum. These descriptors should be robust to the noise and the absence of some of the anatomical structures in one of the images to be registered.


We proposed a novel architecture for medical image registration, which outperforms ANTs and VoxelMorph algorithms for the problem in hand.

\section{Related Work}
\label{sec:related_work}



Deep learning-based medical image registration is still in its nascent stages in contrast to other well-studied tasks such as segmentation or classification. Before the advent of deep learning, most  registration techniques relied on robust feature matching algorithms such as scale-invariant feature transform (SIFT) \cite{crawford2012automated}. The local interest points generated from these algorithms were inherently scale and rotation invariant. The use of these algorithms enabled robust  registration techniques that remain stable against the changes in 3D viewpoint, affine deformations, occlusion, clutter, and noise.

Several successful techniques have been proposed without the use of deep learning. In 2016, Lin et al. \cite{lin2016medical} proposed a registration method based on mimicking the ants foraging for food. Ants algorithm has shown promising results on several medical image registration task in comparison to the other traditional learning techniques. Another well-known method for image registration is the NIFTYReg algorithm which comprises two parts, a global registration followed by a local registration \cite{modat2010lung}. Machado et al ~\cite{machado2018deformable} developed an attribute-matching-based registration method. This method registers intraoperative Ultrasound with preoperative MRI images during image-guided neurosurgery. They utilized Gabor attributes to handle large deformations and absent correspondences.

In the last few years, deep learning has also made its way into image registration. In \cite{hu2018weakly} the authors introduced a weakly supervised CNN for multimodal image registration. The network predicts the displacement field to deform the moving image to match the fixed image, where the labeled images are used during the training and the unlabelled ones are used during the testing. The name weakly supervised derived from the use of anatomical labels to boost the ability of the network to predict the displacement field. The authors proposed this network to tackle the task of registering T2-weighted magnetic resonance images to 3D transrectal ultrasound images from prostate cancer patients. Lee et al.\cite{lin2016medical} proposed image-and-spatial transformer networks (ISTNs) for structure-guided image registration. The proposed ISTNs learn to focus on the structure of interest (SoI). They combined the image transformer network and the spatial transformer network to achieve the registration task. In 2019 Guo proposed a mutual-information-based multi-modal image registration method \cite{guo2019multi}. A year after, Gunnarsson et al. proposed a Laplacian pyramid for medical image registration \cite{gunnarsson2020learning}. The proposed network starts with a rough estimation of the deformation field and refines it in one or more steps. The fixed and the moving images were downsampled at different levels of the pyramid. A similar method is called deep Laplacian Pyramid Image Registration Network (LapIRN) proposed by Tony et al.\cite{mok2020large}. They tried to mimic the traditional multi-resolution-based registration while keeping the non-linearity of the feature maps at different levels of the pyramid.


\section{Proposed method}
\label{sec:proposed_method}

\begin{figure}[htbp]
\includegraphics[width=0.5\textwidth]{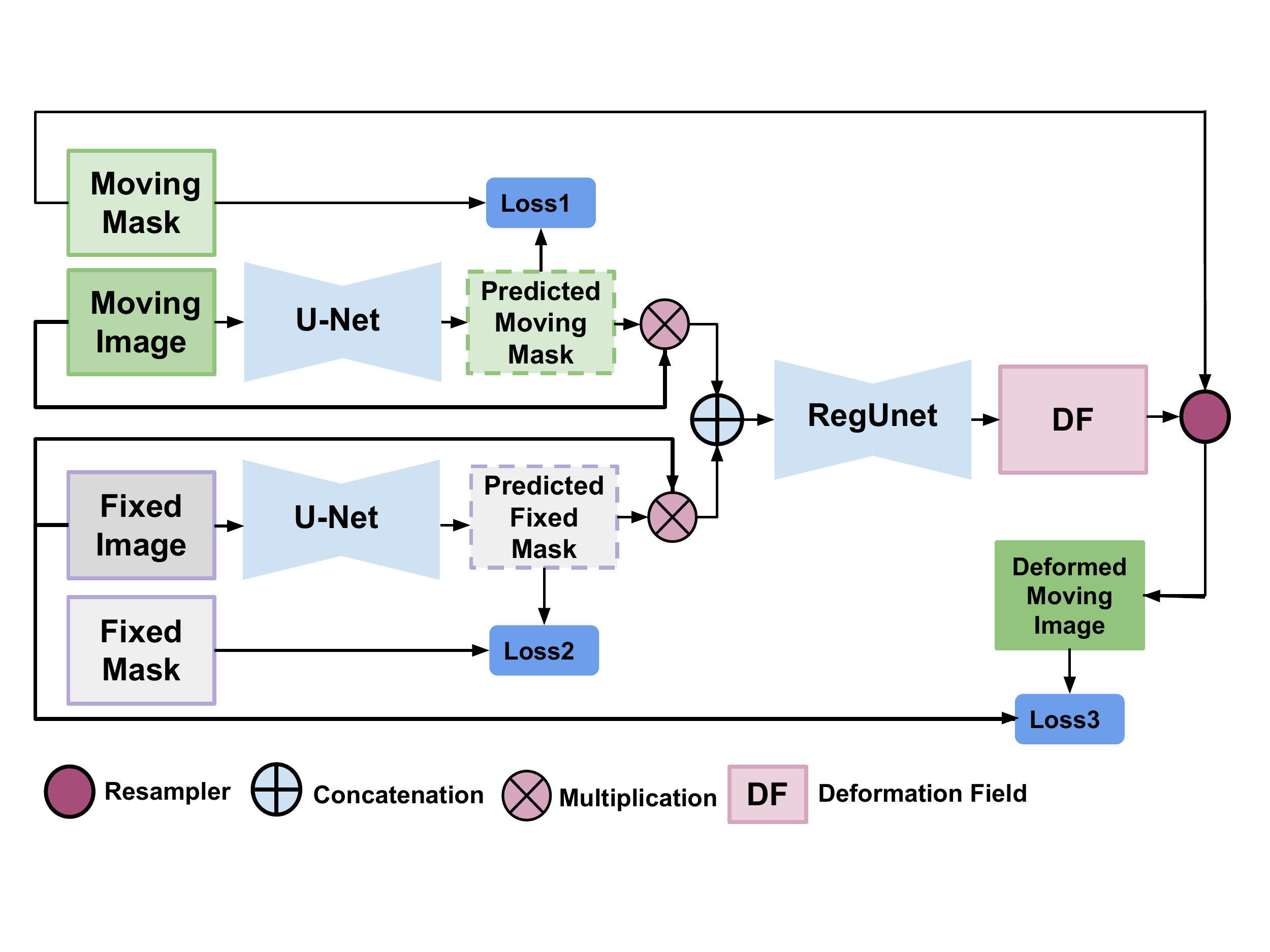}
\caption{The architecture of the proposed method.} 
\label{Arch}
\end{figure}

We propose a weakly supervised method for registering the moving (follow up) scan to the fixed (pre-operative) scan using the landmarks given for every pair of scans. The proposed method consist of  two stages, a segmentation stage, followed by a subsequent registration stage. The segmentation stage comprises of parallel two U-Net architectures that are arranged as shown in figure \ref{Arch}. The first U-Net segments the regions of interest ROIs, small patches of sizes $(9 \times 9 \times 9)$ around the landmarks, of the moving volume.  Similarly, the second U-Net segments the ROIs of the fixed volume. The output of each U-Net is multiplied by the input volume to produce attentive volume. The concatenated outputs of the segmentation networks serve as an input to the registration network. The architecture of the registration network is a U-Net that outputs a deformation field. The deformation field is used for deforming the moving volume to match the fixed one. 

To tackle the problem of class imbalance between the foreground and the background of the segmentation mask, we used the focal loss between the segmentation masks and the predicted segmentation maps \cite{lin2017focal}, as shown in equation \ref{FL}. Here, $\gamma$ is a hyperparameter and we set it to 2 in our experiments. The loss function of the registration network is a combination of two losses, the similarity loss, and the smoothness loss.

\begin{equation}
\label{FL}
    FL(p_t)=-(1-p_t)^{\gamma} log(p_t)
\end{equation} 

The similarity loss is composed of two parts. The first part is the mutual information loss between the Laplacian of Gaussian LoG of the deformed moving volume and the LoG of the fixed volume. While the second part is the local cross-correlation loss between the LoG of the deformed moving volume and the LoG of the fixed volume. The reason behind using the LoG is due to the changes in the appearance of the tissue caused by the existence of the tumor. Thus, choosing a semantic information-based loss might work better than an intensity-based loss.  

The local cross correlation between the fixed volume $f$ and the moving volume $m$ after deforming it by the deformation field $\Phi$ is given by \ref{CC} :
\begin{equation}\label{CC}
cc(f,m\circ \Phi) = \sum_{\boldsymbol{p}\in\boldsymbol{\Omega}} \frac{(\sum_{\boldsymbol{p}_i}(f(\boldsymbol{p}_i-\hat{f}(\boldsymbol{p}))([m \circ \Phi](\boldsymbol{p}_i)-[\hat{m} \circ \Phi](\boldsymbol{p})))^2}{(\sum_{\boldsymbol{p}_i}(f(\boldsymbol{p}_i)-\hat{f}(\boldsymbol{p}))^2)(\sum_{\boldsymbol{p}_i}([m\circ \Phi](\boldsymbol{p}_i)-[\hat{m} \circ \Phi](\boldsymbol{p}))^2)}
\end{equation}

The mutual information between $f$ and ($m\circ \Phi$) is given by \ref{MI}:
\begin{equation}
\label{MI}
    I(f,m\circ \Phi) = \sum_{a\in f, b \in m\circ \Phi}p(a,b) log(\frac{p(a,b)}{p(a)p(b)})
\end{equation}

The smoothness loss is the L2 norm of the Laplacian of the deformation field as in \ref{Smooth}.

\begin{equation}\label{Smooth}
L_{smooth}(\Phi) = \sum_{\boldsymbol{p}\in \Omega}\lVert \triangledown_{\boldsymbol{u}}(\boldsymbol{p})\rVert
\end{equation}
where $\triangledown_{\boldsymbol{u}}(\boldsymbol{p_}) = (\frac{\partial \boldsymbol{u}(\boldsymbol{p})}{\partial x}, \frac{\partial \boldsymbol{u}(\boldsymbol{p})}{\partial y}, \frac{\partial \boldsymbol{u}(\boldsymbol{p})}{\partial z})$, and $\frac{\partial \boldsymbol{u}(\boldsymbol{p})}{\partial x} \approx \boldsymbol{u}(p_x+1,p_y,p_z)-\boldsymbol{u}(p_x,p_y,p_z)$.

The total loss function of the registration network is given by \ref{reg_loss}.
\begin{equation}
    \label{reg_loss}
    Loss = -cc -I + L_{smooth}  
\end{equation}
\section{Data and Experiments}
\label{sec:Data}

The dataset contains pairs of scans -- the pre-operative MRI scans, and the corresponding follow-up MRI scans. Each pair of these scans belongs to the same patient who was cured of glioma. For every patient T1, contrast-enhanced T1-weighted (T1-CE), T2-weighted, and T2 fluid attention inversion recovery (FLAIR) sequences were provided. The training dataset consists of 140 cases and the validation dataset contains 20 cases.

Anatomical markers such as blood vessels bifurcations, the shape of the cortex, and the midline of the brain were used to define the landmarks of the scans of every case. The number of the landmarks differs from one case to another, ranging from 6 to 50 landmarks per case.

We trained our network end to end using three 12GB GPUs. We distributed the U-Net architectures over the GPUs, one on each GPU. All the U-Nets consist of three levels. The number of feature maps starts from 8 and gets doubled when moving from one level to the next one.

First, we trained the network to register a volume (before surgery or a follow-up) to its augmented version; to learn the affine transformation (rotation, scaling, and translation)besides the identity transformation. We used these weights as pre-trained weights of the network trained on registering the follow-up images to the images before the surgery. We used Adam optimizer and step-learning rate scheduler with an initial learning rate of $10^{-4}$ to train our architecture.

\section{Results}
\label{sec:Results}
There were two evaluation metrics. The first metric measured the registration error between the follow-up scan (F) and the preoperative scan (B) for a pair of scans (p) using the median absolute error of the landmarks, which is given by:
\begin{equation}
\label{Absolute Error}
    MAE = Median_{l \in L }(|x_l^B-\hat{x_l}^B|)
\end{equation}
where $x_l^B, x_l^F$ are the coordinates of corresponding landmarks of $l \in L$ the set of landmarks identified in both B and F. 

The second metric measured the robustness of registration. For a pair of scans (p) the robustness is given by the ratio of the successfully registered landmarks to the total number of landmarks in p, which is given by:

 \begin{equation}
   \label{Robustness}  
   R^{B,F}(p)=\frac{|K^{B,F}|}{|L^F|}
 \end{equation}
where $k^{B,F} \subseteq L^F$ are the successfully registered landmarks. The total robustness is given by:
\begin{equation}
\label{totalR}
     R = \frac{1}{P}\sum_{(B,F)\in P}R^{B,F}(p)
\end{equation}

As shown in table \ref{ResultsTable}, the absolute errors of the proposed method are comparable to the ones of ANTs. However, the robustness of the proposed method is higher than the robustness of ANTs. The performance of our proposed method is better than the performance of VoxelMorph for all the evaluation metrics. We reported the average registration time of a pair of volumes on the CPU as we could not find an implementation of ANTs on the GPU. The average registration time of our proposed method on the GPU is less than 6.98 seconds.

\begin{table}[]
\small
\caption{Results of the proposed method versus ANTs and VoxelMorph are reported by the organizers of the challenge on the validation dataset which consists of 20 pairs. AE refers to absolute error. }
\label{ResultsTable}
\begin{tabular}{|c|c|c|c|}

 \hline
 \textbf{Evaluation Metric} & \textbf{ANTs} & \textbf{VoxelMorph}& \textbf{Proposed Method}\\
 \hline
 Median of Median AE & 5.50 & 6.25 & \textbf{5.00}\\ \hline
 Mean of Median AE & \textbf{7.88} & 9.10 & 8.37\\ 
 \hline
 Median of Mean AE & \textbf{5.75} & 6.86 & 6.68 \\
 \hline
 Mean of Mean AE & \textbf{8.55} & 9.29 & 9.09 \\
 \hline
 Mean Robustness & 0.13 & 0.31 & \textbf{0.32} \\
 \hline
 Median Robustness & 0.15 & 0.23 & \textbf{0.28} \\
 \hline
 \vtop{\hbox{\strut Average Running time}\hbox{\strut on CPU in minutes}} & 0.83 & \textbf{0.18} & \textbf{0.18} \\
 \hline
 
\end{tabular}
\end{table}

\section{Conclusion}

We proposed a novel attention-based network for brain MRI registration. Our proposed method surpasses VoxelMorph in terms of median absolute error and robustness. We showed the benefit of training the network to give more attention to the regions around the landmarks in improving the robustness and reducing the registration error.

For future work, we are planning to test the performance of our proposed method on multi-modal image registration problems.

\bibliographystyle{IEEEbib}
\bibliography{strings,refs}
\end{document}